\documentclass[12pt]{article}
\makeatletter
\def\thesection{\arabic{section}}
\@addtoreset{equation}{section}
\def\theequation{\thesection.\arabic{equation}}
\def\appendix{\par
\setcounter{section}{0}
\setcounter{subsection}{0}
\def\thesection{\Alph{section}}}
\makeatother
\makeatletter
\def\abstract#1{\long\def\@abstract{#1}}%
\def\@abstract{}%
\let\@oldmaketitle=\@maketitle%
\def\@maketitle{%
\@oldmaketitle%
\begin{center}\large\bf Abstract\end{center}%
\begin{quotation}\@abstract\end{quotation}%
\vskip 1.5em}%
\makeatother
\makeatletter
\def\eqnarray{%
\stepcounter{equation}%
\let\@currentlabel=\theequation
\global\@eqnswtrue
\global\@eqcnt\z@
\tabskip\@centering
\let\\=\@eqncr
$$\halign to \displaywidth\bgroup\@eqnsel\hskip\@centering
$\displaystyle\tabskip\z@{##}$&\global\@eqcnt\@ne
\hfil$\displaystyle{{}##{}}$\hfil
&\global\@eqcnt\tw@$\displaystyle\tabskip\z@{##}$\hfil
\tabskip\@centering&\llap{##}\tabskip\z@\cr}
\makeatother
\newcommand{\delt}{\Delta t}

\newcommand{\bra}[1]{{\langle{#1}\vert}}
\newcommand{\ket}[1]{{\vert{#1}\rangle}}
\newcommand{\braket}[2]{{\langle{#1}\vert{#2}\rangle}}
\newcommand{\kansu}[2]{{{#1}\!\left({#2}\right)}}
\newcommand{\kakko}[1]{{\left({#1}\right)}}

\newcommand{\wa}[2]{\sum^{#1}_{#2}}
\newcommand{\ffz}{\mbox{\boldmath$z$}}
\begin{document}

\title{\sl Extension of the Barut-Girardello Coherent State\\
        and Path Integral}
\author{
  Kazuyuki FUJII\thanks{e-mail address : fujii@yokohama-cu.ac.jp}
  and 
  Kunio FUNAHASHI\thanks{e-mail address : funahasi@yokohama-cu.ac.jp }
  \\
  Department of Mathematics, Yokohama City University,\\
  Yokohama 236, Japan}
\date{April, 1997}

\abstract{
We extend the Barut-Girardello coherent state for the representation
of $SU(1,1)$ to the coherent state for a representation of 
$U(N,1)$ and construct the measure.
We also construct a path integral formula for some Hamiltonian.
}
\maketitle\thispagestyle{empty}
\newpage

\section{Introduction}

Harmonic oscillator is a fundamental physical system and one of
the few which is solved exactly.
So it has been well studied not only for the system itself but also
for application to other physical systems.
The coherent state is defined as the eigenstate of the annihilation
operator~\cite{GLAUBER}. 
It is a useful tool for study of the harmonic oscillator.
The properties of the coherent state are also well studied.

Extension of the coherent state to various systems has been made.
As for $SU(1,1)$ group, the Perelomov's generalized coherent state
is well-known~\cite{PERELOMOV}.
The treatment is very easy.
Extension of the generalized coherent state to a representation of
$U(N,1)$ has been made~\cite{RON:CPN}.
Making use of the coherent state, we have shown the 
WKB-exactness~\cite{RON:SUTWO,RON:CPN}, which means that the WKB
approximation gives the exact result.
Also we have extended the periodic coherent 
state~\cite{KASHIWA,RON:NR} to the ``multi-periodic coherent state'' 
of a representation
of $U(N,1)$ and have shown the WKB-exactness~\cite{RON:MULTIN,RON:MULTIC}.

There is another coherent state of $SU(1,1)$ which is known as the
Barut-Girardello (BG) coherent state~\cite{BG}.
The BG coherent state is defined as the eigenstate of the lowering operator.
The BG coherent state has a charasteristic feature.
The eigenvalue $K$ of the Casimir operator is valid for 
$K>0$~\cite{BVM}, in sipite of the range of the representation of $SU(1,1)$
is $K\ge1/2$~\cite{WYBOURNE}.

Extension of the BG coherent state has been made to some
groups~\cite{DQ}.
In this paper we extend the BG coherent state to a representation
of $U(N,1)$ to examine the WKB-exactness and et al.

The contents of this paper are as follows.
In Section \ref{sec:cs} we compare the coherent states of the harmonic 
oscillator, the Perelomov's coherent state and the BG coherent state.
We extend the BG coherent state and construct the measure in 
Section \ref{sec:bgcs}.
In Section \ref{sec:pathintegral} we construct a path integral formula 
in terms of the coherent state constructed in Section \ref{sec:bgcs}.
The last section is devoted to the discussions.

\section{Coherent States}\label{sec:cs}

The coherent state of the harmonic oscillator is defined as the
eigenstate of the annihilation operator such that
\begin{equation}
  \label{cs:hoteigi}
  a\ket{z}
  =
  z\ket{z}\ ,\
  \kakko{z\in {\bf C}}\ ,
\end{equation}
where $a(a^\dagger)$ is the annihilation (creation) operator.
Alternative expression of the coherent state is
\begin{equation}
  \label{cs:hoe}
  \ket{z}
  =
  e^{za^\dagger}\ket{0} .
\end{equation}
In the harmonic oscillator, both are equivalent.
The explicit form is
\begin{equation}
  \ket{z}
  =
  \wa{\infty}{n=0}{z^n\over\sqrt{n!}}\ket{n}\ ,
\end{equation}
in the standard notation.
The conditions of the ``coherent state'' in the Klauder's 
sense~\cite{KLAUDER}
are as follows:
\begin{enumerate}
\item continuity: the vector $\ket{l}$ is a strongly continuous
	function of the label $l$,
\item completeness (resolution of unity): 
\begin{equation}
  I
  =
  \int\ket{l}\bra{l}dl\ .
\end{equation}
\end{enumerate}

Now consider the extension to the representation of $SU(1,1)$.
$su(1,1)$ algebra is
\begin{eqnarray}
  &&
  [K_3,K_\pm]
  =
  \pm K_\pm\ ,\
  [K_-,K_+]
  =
  2K_3\ ,
  \nonumber\\
  &&
  K_\pm
  =
  \pm(K_1\pm iK_2)\ ,
\end{eqnarray}
and the representation is
\begin{equation}\label{cs:sudaisu}
  \{
  \ket{K,m}\vert m=0,1,2,\cdots
  \}\ ,\
  K\ge{1\over2}\ ,\
  \textrm{($2K$ is an eigenvalue of the Casimir operator)}\ ,
\end{equation}
which satisfies
\begin{eqnarray}
  K_3\ket{K,m}
  &=&
  (K+m)\ket{K,m}\ ,
  \nonumber\\
  K_+\ket{K,m}
  &=&
  \sqrt{(m+1)(2K+m)}\ket{K,m+1}\ ,
  \nonumber\\
  K_-\ket{K,m}
  &=&
  \sqrt{m(2K+m-1)}\ket{K,m-1}\ .
\end{eqnarray}
The extension of (\ref{cs:hoe}) is known as the Perelomov's 
``generalized coherent state'', whose form is
\begin{eqnarray}
  \label{cs:pcs}
  \ket{\xi}
  \equiv
  e^{\xi K_+}\ket{K,0}
  =
  \wa{\infty}{m=0}\xi^m{2K+m-1\choose m}^{1/2}\ket{K,m}\ ,\ 
  \xi
  &\in&
  \kansu{D}{1,1}\ ,
\end{eqnarray}
where 
\begin{equation}
  \kansu{D}{1,1}
  =
  \big\{
  \xi\in{\bf C}\big\vert\vert\xi\vert<1
  \big\}\ .
\end{equation}
The inner product is
\begin{equation}
  \braket{\xi}{\xi^\prime}
  =
  {1\over\kakko{1-\xi^*\xi^\prime}^{2K}}\ ,
\end{equation}
and the resolution of unity is
\begin{equation}
  \label{cs:pcsrou}
  {2K-1\over\pi}\int_{\kansu{D}{1,1}}
  {d\xi^*d\xi\over\kakko{1-\vert\xi\vert^2}^{-2K+2}}
  \ket{\xi}\bra{\xi}=1_K\ ,
\end{equation}
where
\begin{equation}
  d\xi^*d\xi\equiv\kansu{d}{{\rm Re}\xi}\kansu{d}{{\rm Im}\xi}\ .
\end{equation}

On the other hand, the extension of (\ref{cs:hoteigi}) is known as the Barut-Girardello
coherent state which satisfies
\begin{equation}
  K_-\ket{z}=z\ket{z}\ .
\end{equation}
The explicit form (apart from the normalization factor) is
\begin{equation}
  \label{cs:bgcs}
  \ket{z}=\wa{\infty}{n=0}{z^n\over\sqrt{n!\kakko{2K}_n}}\ket{K,n}\ ,\
z\in{\bf C}\ ,
\end{equation}
where 
\begin{equation}
  (a)_n\equiv a\cdot(a+1)\cdots(a+n-1)\ .
\end{equation}
The inner product is
\begin{equation}
  \braket{z}{z^\prime}
  =
  \kansu{\Gamma}{2K}\kakko{z^*z^\prime}^{-K+{1\over2}}
  \kansu{I_{2K-1}}{2\sqrt{z^*z^\prime}}\ ,
\end{equation}
where $\kansu{\Gamma}{p}$ is the Gamma function:
\begin{equation}
  \kansu{\Gamma}{p}
  =
	\int^\infty_0dte^{-t}t^{p-1}\ ,
\end{equation}
 and $\kansu{I_\nu}{z}$
is the first kind modified Bessel function:
\begin{equation}
  \kansu{I_\nu}{z}
  =
  \kakko{z\over2}^\nu\wa{\infty}{n=0}
  {\kakko{z/2}^{2n}\over n!\kansu{\Gamma}{\nu+n+1}}\ .
\end{equation}
The resolution of unity is
\begin{eqnarray}
  \label{cs:bgcsrou}
  \int\kansu{d\mu}{z,z^*}\ket{z}\bra{z}=1_K\ ,
  \nonumber\\
  \kansu{d\mu}{z,z^*}
  =
  {2\kansu{K_{2K-1}}{2\vert z\vert}\over\pi\kansu{\Gamma}{2K}}
  \vert z\vert^{2K-1}dz^*dz\ .
\end{eqnarray}

In contrast with the coherent state of the harmonic oscillator,
(\ref{cs:pcs}) and (\ref{cs:bgcs}) are not equivalent.
Especially it is remarkable that (\ref{cs:bgcsrou}) holds for $K>0$,
while (\ref{cs:pcsrou}) holds for $K\ge1/2$.

\section{Extension of Barut-Girardello Coherent State}\label{sec:bgcs}

In this section we construct the BG type coherent state for some $U(N,1)$
representation and its measure.

$u(N,1)$ algebra is defined by
\begin{eqnarray}
  \label{ncp:daisuteigi}
  &&
  [E_{\alpha\beta},E_{\gamma\delta}]
  =
  \eta_{\beta\gamma}E_{\alpha\delta}
  -\eta_{\delta\alpha}E_{\gamma\beta}\ ,
  \nonumber\\
  &&
  \eta_{\alpha\beta}
  =
  \kansu{\rm diag}{1,\cdots,1,-1}\ ,\
  \kakko{\alpha,\beta,\gamma,\delta = 1,\cdots,N+1}\ ,
\end{eqnarray}
with a subsidiary condition
\begin{equation}
  -\wa{N}{\alpha=1}E_{\alpha\alpha}+E_{N+1,N+1}=K\ ,\
  \kakko{K=N,N+1,\cdots}\ .
\end{equation}
We identify these generators with creation and annihilation operators
of harmonic oscillators:
\begin{equation}
  \matrix{
    &E_{\alpha\beta}
    =a_\alpha^\dagger a_\beta\ ,&
    E_{\alpha,N+1}
    =a_\alpha^\dagger a_{N+1}^\dagger\ ,
    \cr
    &E_{N+1,\alpha}
    =a_{N+1}a_\alpha\ ,&
    E_{N+1,N+1}
    =a_{N+1}^\dagger a_{N+1}+1\ ,
    \cr
    }
\end{equation}
where $a$, $a^\dagger$ satisfy
\begin{equation}
  [a_\alpha,a_\beta^\dagger]=1\ ,\
  [a_\alpha,a_\beta]=[a_\alpha^\dagger,a_\beta^\dagger]=0\ ,\
  \kakko{\alpha,\beta=1,2,\cdots,N+1}\ .
\end{equation}
The Fock space is
\begin{eqnarray}
  &&
  \left\{
    \ket{n_1,\cdots,n_{N+1}}\vert n_1,n_2,\cdots,n_{N+1}=0,1,2,\cdots
  \right\}\ ,
  \nonumber\\
  \ket{n_1,\cdots,n_{N+1}}
  &\equiv&
  {1\over\sqrt{n_1!\cdots n_{N+1}!}}
  \kakko{a_1^\dagger}^{n_1}\cdots\kakko{a_{N+1}^\dagger}^{n_{N+1}}
  \ket{0,0,\cdots,0}\ ,
  \nonumber\\
  &&
  a_\alpha\ket{0,0,\cdots,0}=0\ ,
\end{eqnarray}
On the representation space it is
\begin{equation}
  1_K
  \equiv
  \wa{\infty}{\left\{ n\right\}=0}
  \ket{n_1,\cdots,n_N,K-1+\sum^N_{\alpha=1}n_\alpha}
  \bra{n_1,\cdots,n_N,K-1+\sum^N_{\alpha=1}n_\alpha}\ ,
\end{equation}
where an abbreviation 
\begin{equation}
  \sum^\infty_{\left\{ n\right\}=0}
  \equiv
  \wa{\infty}{n_1=0}
  \wa{\infty}{n_2=0}
  \cdots
  \wa{\infty}{n_N=0}\ ,
\end{equation}
has been used.

We put the form of the coherent state as
\begin{equation}
  \label{bgcs:teigi}
  \ket{\ffz}
  \equiv
  \wa{\infty}{\left\{ n\right\}=0}
  \kansu{C_{n_1\cdots n_N}}{\ffz}
  z_1^{n_1}\cdots z_N^{n_N}
  \ket{n_1,\cdots,n_N,K-1+\wa{N}{\alpha=1}n_\alpha}\ ,
\end{equation}
to determine the coefficients $\kansu{C_{n_1\cdots n_N}}{\ffz}$'s
so as to satisfy the condition:
\begin{equation}
  \label{bgcs:jouken}
  E_{N+1,\alpha}\ket{\ffz}
  =
  z_\alpha\ket{\ffz}\ ,\
  \kakko{\alpha=1,\ldots,N}\ .
\end{equation}
By noting
\begin{eqnarray}
  a_\alpha\ket{n_1,\ldots,n_\alpha,\ldots,n_{N+1}}
  =
  \sqrt{n_\alpha}
  \ket{n_1,\ldots,n_\alpha-1,\ldots,n_{N+1}}\ ,\\
  \kakko{\alpha=1,\ldots,N+1}\ ,\nonumber
\end{eqnarray}
the explicit form of the left-hand side of (\ref{bgcs:jouken}) is
\begin{eqnarray}
  E_{N+1,\alpha}\ket{\ffz}
  \!\!&=&
  \wa{\infty}{n_1=0}\cdots\wa{\infty}{n_\alpha=0}
  \cdots\wa{\infty}{n_N=0}
  \kansu{C_{n_1\cdots n_N}}{\ffz}
  z_1^{n_1}\cdots z_\alpha^{n_\alpha}\cdots z_N^{n_N}
  \nonumber\\
  &&\times
  \sqrt{n_\alpha}\sqrt{K-1+\wa{N}{\beta=1}n_\beta}
  \ket{n_1,\ldots,n_\alpha-1,\ldots,n_N,K-1+\wa{N}{\beta=1}n_\beta-1}
  \nonumber\\
  &=&
  z_\alpha
  \wa{\infty}{\left\{ n\right\}=0}
  \kansu{C_{n_1\ldots\kakko{n_\alpha+1}\ldots n_N}}{\ffz} 
  \sqrt{n_\alpha+1}\sqrt{K+\wa{N}{\beta=1}n_\beta}
  \nonumber\\
  &&\times
  z_1^{n_1}\ldots z_\alpha^{n_\alpha}\ldots z_N^{n_N}
  \ket{n_1,\ldots,n_\alpha,\ldots,n_N,K-1+\wa{N}{\beta=1}n_\beta}\ ,
\end{eqnarray}
where a shift $n_\alpha\to n_\alpha-1$ has been made in the second equality.
Thus (\ref{bgcs:jouken}) leads to a recursion relation:
\begin{equation}
  \kansu{C_{n_1\ldots \kakko{n_\alpha+1}\ldots n_N}}{\ffz}
  \sqrt{n_\alpha+1}\sqrt{K+\wa{N}{\beta=1}n_\beta}
  =
  \kansu{C_{n_1\ldots n_N}}{\ffz}\ .
\end{equation}
This is easily solved to become
\begin{equation}\label{bgcs:zenka}
  \kansu{C_{n_1\ldots n_\alpha\ldots n_N}}{\ffz} 
  =
  \sqrt{\kakko{K+\wa{N}{\beta=1}n_\beta-n_\alpha-1}!\over
    n_\alpha!\kakko{K+\wa{N}{\beta=1}n_\beta-1}!}
  \kansu{C_{n_1\ldots0\ldots n_N}}{\ffz}\ .
\end{equation}
Application of (\ref{bgcs:zenka}) to all $\alpha$'s 
$\kakko{\alpha=1,\ldots,N}$ leads to
\begin{equation}
  \label{bgcs:zenkatoki}
  \kansu{C_{n_1\ldots n_N}}{\ffz} 
  =
  \sqrt{\kakko{K-1}!\over
    n_1!\ldots n_N!\kakko{K+\wa{N}{\beta=1}n_\beta-1}!}
  \kansu{C_{0\ldots0}}{\ffz}\ .
\end{equation}
Putting (\ref{bgcs:zenkatoki}) into (\ref{bgcs:teigi}),
we obtain the form of the coherent state:
\begin{equation}
  \label{bgcs:kata}
  \ket{\ffz}
  =
  \kansu{C}{\ffz}
  \wa{\infty}{\left\{ n\right\}=0}
  \sqrt{\kakko{K-1}!\over
    n_1!\ldots n_N!\kakko{K+\wa{N}{\beta=1}n_\beta-1}!}
  z_1^{n_1}\ldots z_N^{n_N}
  \ket{n_1,\ldots,n_N,K-1+\wa{N}{\alpha=1}n_\alpha}\ ,
\end{equation}
where we have written $\kansu{C_{0\ldots0}}{\ffz}$ as 
$\kansu{C}{\ffz}$.
$\kansu{C}{\ffz}$ is  a normalization factor and affects no physical
quantity, so hereafter we simply put $\kansu{C}{\ffz}=1$.
Especially, in the $N=1$ case (with $K\to 2K$), (\ref{bgcs:kata})
becomes
\begin{equation}
  \ket{z}
  =
  \wa{\infty}{n=0}
  \sqrt{\kakko{2K-1}!\over n!\kakko{2K+n-1}!}z^n\ket{K,n}\ ,
\end{equation}
which coincides with (\ref{cs:bgcs}), where $\ket{n,2K-1+n}$ has been
identified with $\ket{K,n}$ in the representation of $SU(1,1)$,
(\ref{cs:sudaisu}).

The inner product of the coherent states is
\begin{equation}
  \label{bgcs:naiseki}
  \braket{\ffz}{\ffz^\prime}
  =
  \wa{\infty}{\left\{ n\right\}=0}
  {\kakko{K-1}!\over
    n_1\ldots n_N!\kakko{K+\wa{N}{\alpha=1}n_\alpha-1}!}
  \kakko{z_1^*z_1^\prime}^{n_1}\cdots
  \kakko{z_N^*z_N^\prime}^{n_N}
  =
  \kansu{F_N}{K;\kakko{z_\alpha^*z_\alpha^\prime}}\ ,
\end{equation}
where 
\begin{eqnarray}
  \label{bgcs:fteigi}
  \kansu{F_N}{K;\ffz}
  &\equiv&
  \kansu{F_N}{K;z_1,\ldots,z_N}
  \nonumber\\
  &\equiv&
  \wa{\infty}{\left\{ n\right\}=0}
  {\kakko{K-1}!\over
    n_1!\ldots n_N!\kakko{K+\wa{N}{\alpha=1}n_\alpha-1}!}
  \kakko{z_1}^{n_1}\ldots\kakko{z_N}^{n_N}\ .
\end{eqnarray}
In the $N=1$ case (\ref{bgcs:fteigi}) becomes
\begin{equation}
  \kansu{F_1}{2K;z}
  =
  \kansu{{}_0F_1}{2K;z}
  =
  \kansu{\Gamma}{2K}z^{-K+{1\over2}}\kansu{I_{2K-1}}{2\sqrt{z}}\ ,
\end{equation}
where ${}_0F_1(2K;z)$ is the hypergeometric function and 
$I_\nu(z)$ is the first kind modified Bessel function.

Now we construct the measure so as to satisfy the resolution of unity:
\begin{equation}
  \label{bgcs:taninobunkai}
  \int \kansu{d\mu}{\ffz,\ffz^\dagger}\ket{\ffz}\bra{\ffz}
  =
  1_K\ .
\end{equation}
The explicit form of the left-hand side of (\ref{bgcs:taninobunkai}) is
\begin{eqnarray}
  \label{bgcs:rouarawa}
%  &&
  \int \kansu{d\mu}{\ffz,\ffz^\prime}\ket{\ffz}\bra{\ffz}
%  \nonumber\\
  &=&
  \wa{\infty}{\left\{ n\right\}=0}
  \wa{\infty}{\left\{ m\right\}=0}
  \sqrt{\kakko{K-1}!\over
    n_1!\ldots n_N!\kakko{K+\wa{N}{\alpha=1}n_\alpha-1}!}
  \nonumber\\
  &&\times
  \sqrt{\kakko{K-1}!\over
    m_1!\ldots m_N!\kakko{K+\wa{N}{\alpha=1}m_\alpha-1}!}
  \ket{\left\{ n\right\}}\bra{\left\{ m\right\}}
  \nonumber\\
  &&\times
  \int \kansu{d\mu}{\ffz,\ffz^\dagger}
  z_1^{n_1}\cdots z_N^{n_N}
  \kakko{z_1^*}^{m_1}\cdots\kakko{z_N^*}^{m_N}\ ,
\end{eqnarray}
where 
\begin{eqnarray}
  \kansu{d\mu}{\ffz,\ffz^\dagger}
  &\equiv&
  \kansu{\sigma}{\ffz,\ffz^\dagger}[d\ffz^\dagger d\ffz]\ ,
  \nonumber\\
  {}[d\ffz^\dagger d\ffz]
  &\equiv&
  \prod^N_{\alpha=1}\kansu{d}{{\rm Re}z}\kansu{d}{{\rm Im}z}\ ,
\end{eqnarray}
and an abbreviation
\begin{equation}
  \ket{\left\{ n\right\}}
  \equiv
  \ket{n_1,\cdots,n_N,K-1+\wa{N}{\alpha=1}n_\alpha}\ ,
\end{equation}
has been used.
We put
\begin{equation}
  z_\alpha
  =
  \sqrt{r_\alpha}e^{i\theta_\alpha}\ ,\ 
  \kakko{\alpha=1,\ldots,N}\ ,
\end{equation}
and assume that $\sigma$ depends only on the radial parts:
\begin{eqnarray}
  \kansu{d\mu}{\ffz,\ffz^\dagger}
  =
  \kakko{1\over2}^N
  \kansu{\sigma}{r_1,\ldots,r_N}
  dr_1d\theta_1\cdots dr_Nd\theta_N\ ,
  \nonumber\\
  \kakko{d\ffz^\dagger d\ffz
    =
    \kakko{1\over2}^N
    dr_1d\theta_1\cdots dr_Nd\theta_N}\ .
\end{eqnarray}
Then (\ref{bgcs:rouarawa}) becomes
\begin{eqnarray}
  (\ref{bgcs:rouarawa})
  &=&
  \wa{\infty}{\left\{ n\right\}=0}
  \wa{\infty}{\left\{ m\right\}=0}
  \sqrt{\kakko{K-1}!\over
    n_1!\cdots n_N!\kakko{K+\wa{N}{\alpha=1}n_\alpha-1}!}
  \nonumber\\
  &&
  \times
  \sqrt{\kakko{K-1}!\over
    m_1!\cdots m_N!\kakko{K+\wa{N}{\alpha=1}m_\alpha-1}!}
  \ket{\left\{ n\right\}}\bra{\left\{ m\right\}}
  \nonumber\\
  &&
  \times
  \kakko{1\over2}^N
  \int^\infty_0dr_1\cdots\int^\infty_0dr_N
  \kansu{\sigma}{r_1,\cdots,r_N}
  \int^{2\pi}_0d\theta_1\cdots\int^{2\pi}_0d\theta_N
  \nonumber\\
  &&
  \times
  r_1^{n_1+m_1\over2}\cdots r_N^{n_N+m_N\over2}
  e^{i\kakko{n_1-m_1}\theta_1}\cdots
  e^{i\kakko{n_N-m_N}\theta_N}
  \nonumber\\
  &=&
  \pi^N
  \wa{\infty}{\left\{n\right\}=0}
  {\kakko{K-1}!\over n_1!\cdots n_N!
    \kakko{K+\wa{N}{\alpha=1}n_\alpha-1}!}
  \ket{\left\{ n\right\}}\bra{\left\{ m\right\}}
  \nonumber\\
  &&
  \times
  \int^\infty_0dr_1\cdots\int^\infty_0dr_N
  \kansu{\sigma}{r_1,\cdots,r_n}
  r_1^{n_1}\cdots r_N^{n_N}\ .
\end{eqnarray}
Thus $\kansu{\sigma}{r_1,\cdots,r_N}$ must satisfy
\begin{eqnarray}
  \label{bgcs:mitasu}
  &&\pi^N\kansu{\Gamma}{K}
  \int^\infty_0dr_1\cdots\int^\infty_0dr_N
  \kansu{\sigma}{r_1,\cdots,r_N}
  r_1^{n_1}\cdots r_N^{n_N}
  \nonumber\\
  &&=
  \kansu{\Gamma}{n_1+1}\cdots\kansu{\Gamma}{n_N+1}
  \kansu{\Gamma}{K+\wa{N}{\alpha=1}n_\alpha}\ .
\end{eqnarray}
After some considerations, we found the following formula.
\begin{eqnarray}
  \label{bgcs:koushiki}
  &&\int^\infty_0dr_1 r_1^{s_1}\cdots
  \int^\infty_0dr_N r_N^{s_N}
  2\kakko{r_1+\cdots+r_N}^{K-N\over2}
  \kansu{K_{K-N}}{2\sqrt{r_1+\cdots+r_N}}
  \nonumber\\
  &=&
  \kansu{\Gamma}{s_1+1}\cdots\kansu{\Gamma}{s_N+1}
  \kansu{\Gamma}{K+\wa{N}{\alpha=1}s_\alpha}\ .
\end{eqnarray}
(See appendix \ref{sec:warewarenokoushiki} for the derivation of the formula.)
This is verified as follows.
In the left-hand side of (\ref{bgcs:koushiki}), we put
\begin{eqnarray}
  r_1 &=& \xi_1\kakko{1-\xi_2}\ ,\nonumber\\
  r_2 &=& \xi_1\xi_2\kakko{1-\xi_3}\ ,\nonumber\\
  &\vdots&\nonumber\\
  r_{N-1} &=& \xi_1\xi_2\cdots\xi_{N-1}\kakko{1-\xi_N}\ ,\nonumber\\
  r_N &=& \xi_1\xi_2\cdots\xi_N\ ,
  \nonumber\\
  &&
  \left(
    {\kansu{\partial}{r_1,\cdots,r_N}
      \over
      \kansu{\partial}{\xi_1,\cdots,\xi_N}}
    =
    \xi_1^{N-1}\xi_2^{N-2}\cdots\xi_{N-2}^2\xi_{N-1}
  \right)\ ,
\end{eqnarray}
to obtain
\begin{eqnarray}
  \label{bgcs:hidari}
  \textrm{(l.h.s.)}
  &=&
  2\int^\infty_0d\xi_1 \xi_1^{N-1+s_1+\cdots+s_N+{K-N\over2}}
  \kansu{K_{K-N}}{2\sqrt{\xi_1}}
  \nonumber\\
  &&
  \times
  \int^1_0d\xi_2 \xi_2^{N-2+s_2+\cdots+s_N}
  \kakko{1-\xi_2}^{s_1}
  \int^\infty_0d\xi_3 \xi_3^{N-3+s_3+\cdots+s_N}
  \kakko{1-\xi_3}^{s_2}\cdots
  \nonumber\\
  &&
  \times
  \int^1_0d\xi_{N-1}\xi_{N-1}^{1+s_{N-1}+s_N}
  \kakko{1-\xi_{N-1}}^{s_{N-2}}
  \int^1_0d\xi_N\kakko{1-\xi_N}^{s_{N-1}}\xi_N^{s_N}
  \nonumber\\
  &=&
  2\int^\infty_0d\xi_1 \xi_1^{{K+N\over2}+s_1+\cdots+s_N-1}
  \kansu{K_{K-N}}{2\sqrt{\xi_1}}
  \nonumber\\
  &&
  \times
  \kansu{B}{s_1+1,N-1+s_2+\cdots+s_N}\cdots
  \kansu{B}{s_N+1,s_{N-1}+1}\ ,
\end{eqnarray}
where we have used the integral expression of the Beta function:
\begin{equation}
  \kansu{B}{p,q}
  =
  \int^1_0dt t^{p-1}\kakko{1-t}^{q-1}\ .
\end{equation}
By putting
\begin{equation}
  \sqrt{\xi_1}=u\ ,
\end{equation}
the $\xi_1$-integral in (\ref{bgcs:hidari}) becomes
\begin{eqnarray}
  \textrm{(the $\xi_1$-integral part)}
  &=&
  \int^\infty_0du u^{K+N+2\kakko{s_1+\cdots+s_N}-1}
  \kansu{K_{K-N}}{2u}
  \nonumber\\
  &=&
  {1\over4}
  \kansu{\Gamma}{K+s_1+\cdots+s_N}
  \kansu{\Gamma}{N+s_1+\cdots+s_N}\ ,
\end{eqnarray}
where the formula
\begin{equation}\label{bgcs:iwanamikoushiki}
  \int^\infty_0dx x^{\mu-1}\kansu{K_\nu}{ax}
  =
  {1\over4}
  \kakko{2\over a}^\mu
  \kansu{\Gamma}{\mu+\nu\over2}
  \kansu{\Gamma}{\mu-\nu\over2}\ ,\
  (a>0,\ {\rm Re}\mu>\vert{\rm Re}\nu\vert)\ ,
\end{equation}
has been used (see appendix \ref{sec:iwanamikoushiki}).
By noting the relation
\begin{equation}
  \kansu{B}{p,q}
  =
  {\kansu{\Gamma}{p}\kansu{\Gamma}{q}\over\kansu{\Gamma}{p+q}}\ ,
\end{equation}
(\ref{bgcs:hidari}) finally becomes
\begin{equation}
  (\ref{bgcs:hidari})
  =
  \kansu{\Gamma}{s_1+1}\cdots\kansu{\Gamma}{s_N+1}
  \kansu{\Gamma}{K+s_1+\cdots+s_N}\ ,
\end{equation}
which is just the right-hand side of (\ref{bgcs:koushiki}).

Comparing (\ref{bgcs:mitasu}) with (\ref{bgcs:koushiki}), we find that
\begin{equation}
  \kansu{\sigma}{r_1,\cdots,r_N}
  =
  {2\over\pi^N\kansu{\Gamma}{K}}
  \kakko{r_1+\cdots+r_N}^{K-N\over2}
  \kansu{K_{K-N}}{2\sqrt{r_1+\cdots+r_N}}\ .
\end{equation}
Therefore we obtain the measure:
\begin{equation}\label{bgcs:measure}
  \kansu{d\mu}{\ffz,\ffz^\dagger}
  =
  {2\|\ffz\|^{K-N}\kansu{K_{K-N}}{2\|\ffz\|}\over\pi^N\kansu{\Gamma}{K}}
  [d\ffz^\dagger d\ffz]\ ,
\end{equation}
where 
\begin{equation}
  \|\ffz\|\equiv\sqrt{\ffz^\dagger\ffz}\ .
\end{equation}
In the $N=1$ case, (\ref{bgcs:measure}) ($K\to 2K$) is
\begin{equation}
  \kansu{d\mu}{z,z^*}
  =
  {2\vert z\vert^{2K-1}\kansu{K_{2K-1}}{2\vert z\vert}
    \over
    \pi\kansu{\Gamma}{2K}}
  [dz^*dz]\ ,
\end{equation}
which is just the measure of BG coherent state (\ref{cs:bgcsrou}).
The explicit form of (\ref{bgcs:taninobunkai}) is
\begin{eqnarray}
  \label{bgcs:gutai}
  &&
  \int{2\|\ffz\|^{K-N}\kansu{K_{K-N}}{2\|\ffz\|}
    \over
    \pi^2\kansu{\Gamma}{K}}
  [d\ffz^\dagger d\ffz]
  \nonumber\\
  &&
  \times
  \wa{\infty}{\left\{ n\right\}=0}
  z_1^{n_1}\cdots z_N^{n_N}
  \ket{n_1,\cdots,n_N,K-1+\wa{N}{\alpha=1}n_\alpha}
  \nonumber\\
  &&
  \times
  \wa{\infty}{\left\{ m\right\}=0}
  \kakko{z_1^*}^{m_1}\cdots\kakko{z_N^*}^{m_N}
  \bra{m_1,\cdots,m_N,K-1+\wa{N}{\alpha=1}m_\alpha}
  =
  {\bf 1}_K\ .
\end{eqnarray}
{\em It is remarkable that (\ref{bgcs:gutai}) holds for $K>0$.}

\section{Construction of Path Integral Formula}
\label{sec:pathintegral}

We construct a path integral formula with a Hamiltonian
\begin{equation}
  \hat H
  \equiv
  \wa{N+1}{\alpha=1}c_\alpha E_{\alpha\alpha}
  =
  \wa{N}{\alpha=1}\mu_\alpha E_{\alpha\alpha}+Kc_{N+1}{\bf 1}_K\ ,\
  \kakko{\mu_\alpha\equiv c_\alpha+c_{N+1}}\ .
\end{equation}
We have shown the WKB-exactness of the trace formula 
with this Hamiltonian in terms of the 
``generalized coherent state''~\cite{RON:CPN} and the ``multi-periodic
coherent state''~\cite{RON:MULTIN}.
In this section we write down the trace formula in terms of our
BG coherent state.

The matrix element of the Hamiltonian is 
\begin{equation}
  \bra{\ffz}{\hat H}\ket{\ffz^\prime}
  =
  Kc_{N+1}\kansu{F_N}{K;\left( z_\alpha^* z_\alpha^\prime\right)}
  +\wa{N}{\alpha=1}\mu_\alpha z_\alpha^*z_\alpha^\prime
  \kansu{F_N}{K+1;\left( z_\alpha^* z_\alpha^\prime\right)}\ ,
\end{equation}
where $\kansu{F_N}{K;\ffz}$ is given in (\ref{bgcs:fteigi}).

The Feynman kernel is defined by
\begin{equation}
  \kansu{K}{\ffz_F,\ffz_I;T}
  \equiv
  \bra{\ffz_F}e^{-i\hat HT}\ket{\ffz_I}
  =
  \lim_{M\to\infty}
  \bra{\ffz_F}\kakko{1-i\delt\hat H}^M\ket{\ffz_I}\ ,\
  \kakko{\delt\equiv T/M}\ ,
\end{equation}
where $\ffz_I(\ffz_F)$ is the initial (final) state and $T$ is time
interval.
The explicit form of the kernel is
\begin{eqnarray}
  \kansu{K}{\ffz_F,\ffz_I;T}
  &=&
  \lim_{M\to\infty}
  \int\prod^{M-1}_{i=1}
  \kansu{d\mu}{\kansu{\ffz}{i},\kansu{\ffz^\dagger}{i}}
  \prod^M_{j=1}
  \bra{\kansu{\ffz}{j}}\kakko{1-i\delt\hat H}\ket{\kansu{\ffz}{j-1}}
  \Bigg\vert^{\kansu{\ffz}{M}=\ffz_F}_{\kansu{\ffz}{0}=\ffz_I}
  \nonumber\\
  &=&
  \lim_{M\to\infty}
  \int\prod^{M-1}_{i=1}
  \kansu{d\mu}{\kansu{\ffz}{i},\kansu{\ffz^\dagger}{i}}
  \prod^M_{j=1}
  \Bigg\{
  \kansu{F_N}{K;\kakko{\kansu{z_\alpha^*}{j}\kansu{z_\alpha}{j-1}}}
  \nonumber\\
  &&
  \times
  \bigg[
  1-i\delt\bigg\{
  Kc_{N+1}
  \nonumber\\
  &&
  +\wa{N}{\alpha=1}\mu_\alpha\kansu{z_\alpha^*}{j}\kansu{z_\alpha}{j-1}
  {\kansu{F_N}{K+1;\kakko{\kansu{z_\alpha^*}{j}\kansu{z_\alpha}{j-1}}}
    \over
    \kansu{F_N}{K;\kakko{\kansu{z_\alpha^*}{j}\kansu{z_\alpha}{j-1}}}
  }
  \bigg\}
  \bigg]
  \Bigg\}
  \nonumber\\
  &=&
  \lim_{M\to\infty}
  \int\prod^{M-1}_{i=1}
  \kansu{d\mu}{\kansu{\ffz}{i},\kansu{\ffz^\dagger}{i}}
  \prod^M_{j=1}
  \Bigg\{
  \kansu{F_N}{K;\kakko{\kansu{z_\alpha^*}{j}\kansu{z_\alpha}{j-1}}}
  \Bigg\}
  \nonumber\\
  &&
  \times
  \exp\bigg[
        -i\delt\wa{M}{k=1}\bigg\{
          Kc_{N+1}
          \nonumber\\
          &&
          +\wa{N}{\alpha=1}\mu_\alpha
          \kansu{z_\alpha^*}{k}\kansu{z_\alpha}{k-1}
          {\kansu{F_N}
            {K+1;\kakko{\kansu{z_\alpha^*}{k}\kansu{z_\alpha}{k-1}}}
          \over
          \kansu{F_N}
            {K;\kakko{\kansu{z_\alpha^*}{k}\kansu{z_\alpha}{k-1}}}}
        \bigg\}
  \bigg]\ ,
\end{eqnarray}
where the resolution of unity (\ref{bgcs:taninobunkai}) has been
inserted in the first equality and $O((\delt)^2)$ terms,
which finally vanish in $M\to\infty$ limit, have been omitted
in the last equality.

The trace formula is defined by
\begin{equation}
  Z
  \equiv
  \int\kansu{d\mu}{\ffz,\ffz^\dagger}
  \bra{\ffz}e^{-i\hat HT}\ket{\ffz}
  =
  \int\kansu{d\mu}{\ffz,\ffz^\dagger}
  \kansu{K}{\ffz,\ffz;T}\ .
\end{equation}
The explicit form is
\begin{eqnarray}\label{pi:gutai}
  Z
  &=&
  \lim_{M\to\infty}
  \int\prod^M_{i=1}
  \kansu{d\mu}{\kansu{\ffz}{i}\kansu{\ffz^\dagger}{i}}
  \prod^M_{j=1}
        \bigg\{
          \kansu{F_N}{K;\kakko{\kansu{z_\alpha^*}{j}
                \kansu{z_\alpha}{j-1}}}
        \bigg\}
  \nonumber\\
  &&
  \times
  \exp\Bigg[
        -i\delt\wa{M}{k=1}\bigg\{
          Kc_{N+1}
  \nonumber\\
  &&
  +\wa{N}{\alpha=1}\mu_\alpha
        \kansu{z_\alpha^*}{k}\kansu{z_\alpha}{k-1}
        {\kansu{F_N}{K+1;\kakko{
          \kansu{z_\alpha^*}{k}\kansu{z_\alpha}{k-1}}}
        \over
        \kansu{F_N}{K;\kakko{
          \kansu{z_\alpha^*}{k}\kansu{z_\alpha}{k-1}}}}
        \bigg\}
  \Bigg]\Bigg\vert_{\kansu{\ffz}{M}=\kansu{\ffz}{0}}
  \nonumber\\
  &=&
  \lim_{M\to\infty}
  e^{-iKc_{N+1}T}
  \int\prod^M_{i=1}
  \kansu{d\mu}{\kansu{\ffz}{i},\kansu{\ffz^\dagger}{i}}
  \prod^M_{j=1}
  \bigg\{
        \kansu{F_N}{K;\kakko{
          \kansu{z_\alpha^*}{j}\kansu{z_\alpha}{j-1}}}
  \bigg\}
  \nonumber\\
  &&
  \times
  \exp\Bigg[
        -i\delt\wa{M}{k=1}\wa{N}{\alpha=1}\mu_\alpha
        \kansu{z_\alpha^*}{k}\kansu{z_\alpha}{k-1}
        {\kansu{F_N}{K+1;\kakko{
          \kansu{z_\alpha^*}{k}\kansu{z_\alpha}{k-1}}}
        \over
        \kansu{F_N}{K;\kakko{
          \kansu{z_\alpha^*}{k}\kansu{z_\alpha}{k-1}}}}
  \Bigg]\ .
\end{eqnarray}
In the $N=1$ case, (\ref{pi:gutai}) is
\begin{eqnarray}
  Z
  &=&
  \lim_{M\to\infty}e^{-ihKT}
  \int\prod^M_{i=1}
  \kakko{{2\over\pi}\kansu{K_{2K-1}}{2\vert\kansu{z}{i}\vert}
  \kansu{I_{2K-1}}{2\sqrt{\kansu{z^*}{i}\kansu{z}{i-1}}}
  [\kansu{dz^*}{i}\kansu{dz}{i}]}
  \nonumber\\
  &&
  \times
  \exp\Bigg[
  -ih\delt\wa{M}{k=1}\sqrt{\kansu{z^*}{k}\kansu{z}{k-1}}
  {\kansu{I_{2K}}{2\sqrt{\kansu{z^*}{k}\kansu{z}{k-1}}}
      \over
      \kansu{I_{2K-1}}{2\sqrt{\kansu{z^*}{k}\kansu{z}{k-1}}}}
    \Bigg]\ ,
\end{eqnarray}
where $h\equiv c_1+c_2$.\\
Even in the $N=1$ case, it seems to be complicated not only
to make the WKB approximation but also to calculate exactly.

\section{Discussion}
\label{sec:giron}

We have extended the BG coherent state for the representation of $SU(1,1)$
to some representation of $U(N,1)$ and constructed the measure.
The eigenvalue of the Casimir operator 
($I=\wa{N+1}{\alpha=1}E_{\alpha\alpha}$), $K$, can be enlarged to $K>0$.

We have also constructed the path integral formula with the same 
Hamiltonian with which the WKB-exactness is examined in terms of some
coherent states.
However, in this case, the form of the coherent state is so complicated
that the WKB approximation as well as the exact calculation seems not
to be easy.
Showing the WKB-exactness is now under consideration.

\appendix
\vspace{15mm}
\begin{center}
  \LARGE\bfseries Appendix
\end{center}

\section{Derivation of the Formula (\protect\ref{bgcs:koushiki})}
\label{sec:warewarenokoushiki}

The definition of the Gamma function immediately leads to the relation:
\begin{equation}
  \label{sec:b:sugu}
  \int^\infty_0dr_\alpha
  e^{-a_\alpha r_\alpha}
  r_\alpha^{s_\alpha}
  =
  a_\alpha^{-\kakko{s_\alpha+1}}
  \kansu{\Gamma}{s_\alpha+1}\ .
\end{equation}
Multiplying equations (\ref{sec:b:sugu}) from $\alpha=1$ to $N$
and putting $a_\alpha=1/x$ for all $\alpha(=1,\cdots,N)$, 
we obtain
\begin{equation}
  \int^\infty_0dr_1 r_1^{s_1}\cdots
  \int^\infty_0dr_N r_N^{s_N}
  e^{-R/x}
  =
  x^{s_1+\cdots+s_N+N}
  \kansu{\Gamma}{s_1+1}\cdots\kansu{\Gamma}{s_N+1}\ ,
\end{equation}
where 
\begin{equation}
  R
  \equiv
  r_1+\cdots+r_N\ .
\end{equation}
In both sides, by multiplying $e^{-x}x^{K-N-1}$ and
by integrating over $x$, the right-hand side becomes
\begin{equation}
  \label{b:migi}
  \textrm{(r.h.s.)}
  =
  \kansu{\Gamma}{s_1+1}\cdots\kansu{\Gamma}{s_N+1}
  \kansu{\Gamma}{K+\wa{N}{\alpha=1}s_\alpha}\ .
\end{equation}
Then the left-hand side is
\begin{eqnarray}
  \label{b:hidari}
  \textrm{(l.h.s)}
  &=&
  \int^\infty_0dr_1 r_1^{s_1}\cdots
  \int^\infty_0dr_N r_N^{s_N}
  \int^\infty_0dx x^{K-N-1}
  e^{-R/x-x}
  \nonumber\\
  &=&
  \int^\infty_0dr_1 r_1^{s_1}\cdots
  \int^\infty_0dr_N r_N^{s_N}
  R^{K-N}
  \int^\infty_0du u^{-K+N-1}
  e^{-u-R/u}\ ,
\end{eqnarray}
where a change of variable, $u=R/x$, has been made 
in the second equality.
By the integral expression of the modified Bessel function:
\begin{equation}
  \kansu{K_\nu}{z}
  =
  {1\over2}
  \kakko{z\over2}^\nu
  \int^\infty_0dt
  t^{-\nu-1}
  e^{-t-z^2/4t}\ ,
\end{equation}
the $u$-integral is written by
\begin{equation}
  \int^\infty_0du u^{-K+N-1}e^{-u-R/u}
  =
  2R^{-{K-N\over2}}\kansu{K_{K-N}}{2\sqrt{R}}\ .
\end{equation}
Thus (\ref{b:hidari}) becomes
\begin{equation}
  \label{b:kekkyoku}
  \textrm{(\ref{b:hidari})}
  =
  \int^\infty_0dr_1 r_1^{s_1}\cdots
  \int^\infty_0dr_N r_N^{s_N}
  2R^{K-N\over2}\kansu{K_{K-N}}{2\sqrt{R}}\ .
\end{equation}
Finally, by (\ref{b:migi}) and (\ref{b:kekkyoku}), we obtain
\begin{eqnarray}
  &&\int^\infty_0dr_1 r_1^{s_1}\cdots
  \int^\infty_0dr_N r_N^{s_N}
  2\kakko{r_1+\cdots+r_N}^{K-N\over2}
  \kansu{K_{K-N}}{2\sqrt{r_1+\cdots+r_N}}
  \nonumber\\
  &=&
  \kansu{\Gamma}{s_1+1}\cdots\kansu{\Gamma}{s_N+1}
  \kansu{\Gamma}{K+\wa{N}{\alpha=1}s_\alpha}\ .
\end{eqnarray}

\section{The Proof of the Formula (\protect\ref{bgcs:iwanamikoushiki})}
\label{sec:iwanamikoushiki}

By the integral representation of the modified Bessel function:
\begin{equation}
  \kansu{K_\nu}{z}
  =
  {\sqrt{\pi}\over\kakko{\nu-{1\over2}}!}
  \kakko{{z\over2}}^\nu
  \int^\infty_1dy
  e^{-zy}
  \kakko{y^2-1}^{\nu-{1\over2}}\ ,\ 
  \textrm{for $\nu>-1/2$}\ ,
\end{equation}
the left-hand side of the formula (\ref{bgcs:iwanamikoushiki}) becomes
\begin{equation}\label{sec:app:a:naru}
  \int^\infty_0dx x^{\mu-1}\kansu{K_\nu}{ax}
  =
  {\sqrt{\pi}\over\kakko{\nu-{1\over2}}!}
  \int^\infty_1dy
  \kakko{y^2-1}^{\nu-{1\over2}}
  \left\{
    \int^\infty_0dx x^{\mu-1}
    \kakko{ax\over2}^\nu
    e^{-axy}
  \right\}\ .
\end{equation}
Changing a variable $x$ to $t$ such that
\begin{equation}
  axy=t\ ,
\end{equation}
leads the $x$-integral to
\begin{eqnarray}
  \int^\infty_0dx x^{\mu-1}
  \kakko{{ax\over2}}^\nu
  e^{-axy}
  =
  {1\over a^\mu y^{\mu+\nu}2^\nu}
  \int^\infty_0dte^{-t}t^{\mu+\nu-1}
  =
  {\kansu{\Gamma}{\mu+\nu}\over a^\mu 2^\nu y^{\mu+\nu}}\ .
\end{eqnarray}
Thus (\ref{sec:app:a:naru}) becomes
\begin{equation}\label{sec:a:tochuu}
  \kakko{\ref{sec:app:a:naru}}
  =
  {\sqrt{\pi}\over\kakko{\nu-{1\over2}}!}
  {\kansu{\Gamma}{\mu+\nu}\over a^\mu2^\nu}
  \int^\infty_1dy{1\over y^{\mu+\nu}}
  \kakko{y^2-1}^{\nu-{1\over2}}\ .
\end{equation}
Further we make a change of a variable such that
\begin{equation}
  y=1/\sqrt{t}\ ,
\end{equation}
to obtain
\begin{eqnarray}\label{sec:a:chuuto}
  \kakko{\ref{sec:a:tochuu}}
  &=&
  {\sqrt{\pi}\over\kakko{\nu-{1\over2}}!}
  {\kansu{\Gamma}{\mu+\nu}\over a^\mu 2^\nu}
  {1\over2}
  \int^1_0dt t^{{\mu-\nu\over2}-1}
  \kakko{1-t}^{\nu+{1\over2}-1}
  \nonumber\\
  &=&
  {\sqrt{\pi}\over\kakko{\nu-{1\over2}}!}
  {\kansu{\Gamma}{\mu+\nu}\over a^\mu 2^{\nu+1}}
  \kansu{B}{{\mu-\nu\over2},{\mu+\nu\over2}}
  \nonumber\\
  &=&
  {\sqrt{\pi}\over a^\mu 2^{\nu+1}}
  \kansu{\Gamma}{\mu-\nu\over2}
  \kansu{\Gamma}{\mu+\nu\over2}
  {\kansu{\Gamma}{\mu+\nu}\over\kansu{\Gamma}{{\mu+\nu\over2}}
    \kansu{\Gamma}{{\mu+\nu\over2}+{1\over2}}} \ ,
\end{eqnarray}
where we have used the integral expression of the Beta function:
\begin{equation}
  \kansu{B}{p,q}
  =
  \int^1_0dt t^{p-1}\kakko{1-t}^{q-1}\ ,
\end{equation}
and the relation:
\begin{equation}
  \kansu{B}{p,q}
  =
  {\kansu{\Gamma}{p}\kansu{\Gamma}{q}\over
    \kansu{\Gamma}{p+q}}\ .
\end{equation}
By the formula:
\begin{equation}
  z!\kakko{z+{1\over2}}!
  =
  2^{-2z-1}\sqrt{\pi}\kakko{2z+1}!\ ,
\end{equation}
the product of the Gamma functions is written as
\begin{equation}
  \kansu{\Gamma}{{\mu+\nu\over2}}
  \kansu{\Gamma}{{\mu+\nu\over2}+{1\over2}}
  =
  2^{-\kakko{\mu+\nu}+1}
  \sqrt{\pi}
  \kansu{\Gamma}{\mu+\nu}\ .
\end{equation}
Thus (\ref{sec:a:chuuto}) becomes
\begin{equation}
  (\ref{sec:a:chuuto})
  =
  {1\over4}
  \kakko{{2\over a}}^\mu
  \kansu{\Gamma}{{\mu-\nu\over2}}
  \kansu{\Gamma}{{\mu+\nu\over2}}\ ,
\end{equation}
which is just the right-hand side of the formula.

\end{document}